\documentclass{article}
\usepackage{spconf,amsmath,graphicx}
\usepackage{enumitem}
\setlist{nosep, leftmargin=14pt}
\usepackage{booktabs} 
\usepackage{amssymb}
\usepackage{color}
\usepackage[colorlinks, urlcolor=black, linkcolor=blue, citecolor=blue]{hyperref}

\title{Multi-Modality Transrectal Ultrasound Video Classification for\\Identification of Clinically Significant Prostate Cancer}
\name{Hong Wu$^{1,2,\dagger}$\thanks{$\dagger$ Hong Wu and Juan Fu contribute equally to this work.}, Juan Fu$^{3,\dagger}$, Hongsheng Ye$^{3}$, Yuming Zhong$^{1,2}$, Xuebin Zou$^{3}$, Jianhua Zhou$^{3,*}$, Yi Wang$^{1,2,*}$\thanks{$*$ Corresponding authors: Jianhua Zhou and Yi Wang.}
\thanks{This work was supported in part by the National Natural Science Foundation of China under Grants 81971631 and 62071305, in part by the Guangdong-Hong Kong Joint Funding for Technology and Innovation under Grant 2023A0505010021, and in part by the Guangdong Basic and Applied Basic Research Foundation under Grant 2022A1515011241.}
}
\address{$^{1}$National-Regional Key Technology Engineering Laboratory for Medical Ultrasound,\\
	Guangdong Key Laboratory for Biomedical Measurements and Ultrasound Imaging,\\
	School of Biomedical Engineering,
	Shenzhen University, Shenzhen, China\\
	$^{2}$Smart Medical Imaging, Learning and Engineering (SMILE) Lab,\\
	Medical UltraSound Image Computing (MUSIC) Lab, Shenzhen, China\\
	$^{3}$ The Department of Ultrasound, State Key Laboratory of Oncology in South China,\\
	Guangdong Provincial Clinical Research Center for Cancer,\\
	Sun Yat-sen University Cancer Center, Guangzhou, China\\}
\begin{document}
%
\maketitle
\begin{abstract}
Prostate cancer is the most common noncutaneous cancer in the world.
Recently, multi-modality transrectal ultrasound (TRUS) has increasingly become an effective tool for the guidance of prostate biopsies.
With the aim of effectively identifying prostate cancer,
we propose a framework for the classification of clinically significant prostate cancer (csPCa) from multi-modality TRUS videos.
The framework utilizes two 3D ResNet-50 models to extract features from B-mode images and shear wave elastography images, respectively.
An adaptive spatial fusion module is introduced to aggregate two modalities' features.
An orthogonal regularized loss is further used to mitigate feature redundancy.
The proposed framework is evaluated on an in-house dataset containing 512 TRUS videos,
and achieves favorable performance in identifying csPCa with an area under curve (AUC) of 0.84.
Furthermore, the visualized class activation mapping (CAM) images generated from the proposed framework may provide valuable guidance for the localization of csPCa, thus facilitating the TRUS-guided targeted biopsy.
Our code is publicly available at~\textit{https://github.com/2313595986/ProstateTRUS}.
\end{abstract}
\begin{keywords}
Clinically significant prostate cancer (csPCa), 
transrectal ultrasound, 
shear wave elastography,
deep learning,
class activation mapping
\end{keywords}
\section{Introduction}
\label{sec:intro}
Prostate cancer (PCa) is the most prevalent cancer among males in the world~\cite{siegel2023cancer}.
Prostate biopsy, the gold standard for PCa diagnosis, has been widely used in clinical practice.
Specifically, PCa can be classified into clinically significant PCa (csPCa, Gleason score $\geq$ 3+4=7) and clinically insignificant PCa (cisPCa, Gleason score $\leq$ 3+3=6)~\cite{matoso2019defining}.
Since csPCa has a worse prognosis compared with cisPCa,
early detection and identification of patients with csPCa is beneficial to improve the survival rate and prognosis.

\begin{figure}[t]
	\centering
	\includegraphics[width=\linewidth]{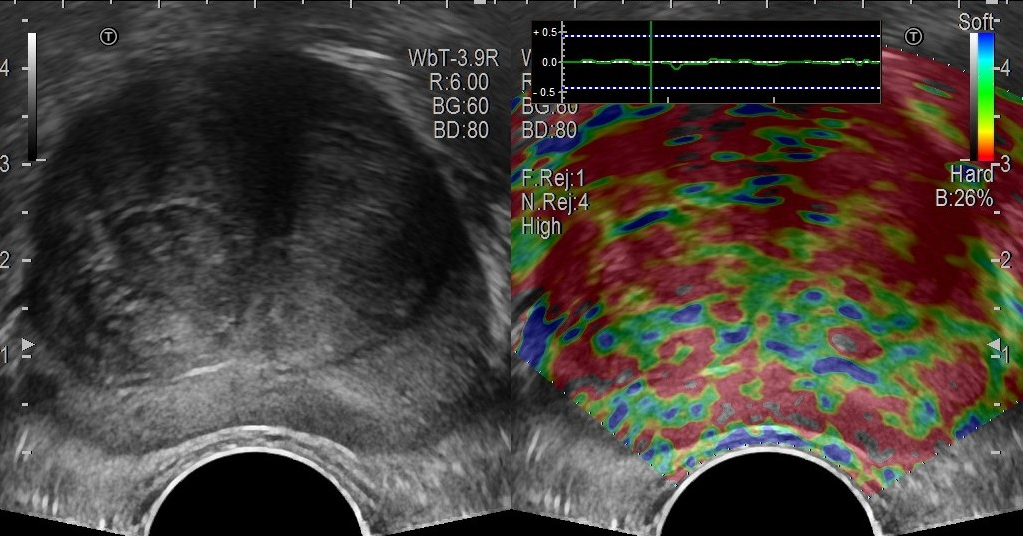}
	\caption{A frame extracted from the transrectal ultrasound (TRUS) video, displaying both B-mode and shear wave elastography (SWE) images.
	The B-mode provides anatomical details, while the SWE depicts tissue stiffness.}
	\label{fig:SWE}
\end{figure}

\begin{figure*}[t]
	\centering
	\includegraphics[width=0.95\textwidth]{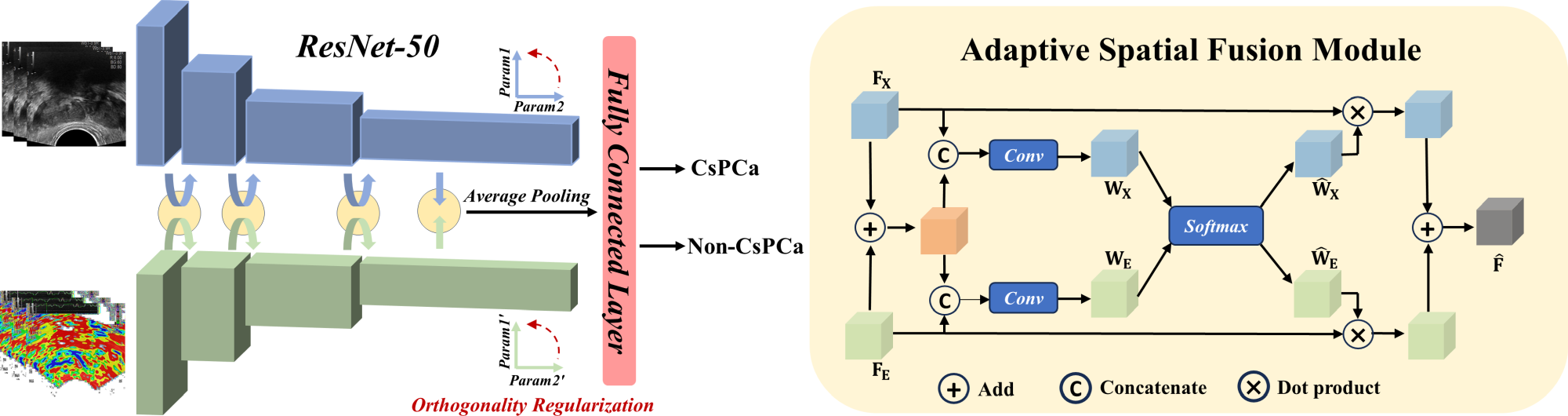}
	\caption{A schematic overview of our proposed framework for csPCa classification in TRUS videos.
		The classification framework consists of two 3D ResNet-50 models to extract features from B-mode images and SWE images, respectively.
		The adaptive spatial fusion module is designed to aggregate two modalities' features.
		The orthogonal regularization encourages weights to be orthogonal, thereby mitigating feature redundancy.
	}
	\label{fig:method}
\end{figure*}

Current guidelines recommend multiparametric magnetic resonance imaging (mp-MRI) as the primary tool for prostate biopsy
due to its capability to depict the localization and morphology of PCa, 
thereby enabling the implementation of targeted biopsy~\cite{mottet2021eau}.
However, some inherent factors restrict the widespread use of MRI-guided targeted biopsies,
such as complex operations, low availability and high cost.
Hence, transrectal ultrasound (TRUS) is a more universally available tool for PCa diagnosis following mp-MRI~\cite{schoots2015magnetic}.
TRUS has the advantages of convenient and real-time imaging,
but the low contrast between malignant and normal tissue reduces its ability to identify PCa.
Thus, TRUS is used mostly with systematic biopsy,
necessitating more puncture cores to increase PCa detection rates~\cite{brown2015recent}.
Meanwhile, it will increase the pain and burden of patients and cause a great waste of medical resources.
Therefore, accurate identification of the target lesion in TRUS during biopsy procedure has long been an actively explored research issue.

Multimodal technological upgrades such as transrectal shear wave elastography (SWE),
as shown in Fig.~\ref{fig:SWE},
have been developed to facilitate TRUS-guided targeted biopsy~\cite{ahmad2013transrectal}.
SWE can be used to qualitatively and quantitatively analyze prostate tissue stiffness.
Clinical studies have demonstrated that the target lesion within the prostate typically exhibits greater stiffness compared to the surrounding normal tissue~\cite{mongiat2019extracellular}.
This can be attributed to the increased proportion of extracellular matrix proteins associated with PCa formation.
Thus, SWE enables the identification and characterization of abnormally stiff regions in the prostate, 
thereby facilitating the TRUS-guided targeted biopsy.

To better support the TRUS-guided targeted biopsy, several algorithms have been developed to help predict csPCa based on TRUS images.
Liang~\textit{et al.}~\cite{liang2021nomogram} employed a multiparametric ultrasound-based radiomics model,
which incorporated B-mode and SWE to predict the csPCa.
This model achieved an area under curve (AUC) of 0.85 on 112 patients.
Wildeboer~\textit{et al.}~\cite{wildeboer2020automated} combined the information from B-mode, SWE and contrast-enhanced ultrasound to develop a classifier for the identification of csPCa.
They achieved an AUC of 0.90 on 48 patients.
However, these studies focused on the analysis of hand-crafted features, and most of them required manual interactions.
Sun~\textit{et al.}~\cite{sun2023three} proposed a mask-guided hierarchical framework for identifying csPCa using B-mode videos.
They first introduced a 2D segmentation network to remove surrounding tissues around the prostate gland.
Subsequently, they constructed classification network on a dataset containing 699 patients' B-mode videos,
and yielded an AUC of 0.86.

The primary motivation of this study is to propose a modality-fusion video classification network for the identification of csPCa.
Our approach utilizes TRUS videos, incorporating B-mode and SWE information, to train the network.
Specifically, we introduce an adaptive spatial fusion module to aggregate these two modalities' features and facilitate the final classification.
During training, an orthogonal regularization strategy is further used to encourage weights to be orthogonal, thereby mitigating feature redundancy.
Additionally, we employ the class activation mapping (CAM) to visualize the indicative areas related to the prediction of csPCa,
which enables the feasibility of TRUS-guided biopsy.

\section{Method}
An overview of our proposed framework for csPCa classification is illustrated in Fig.~\ref{fig:method}.
The framework incorporates two 3D ResNet-50 models~\cite{he2016deep} for feature extraction from the B-mode images and the corresponding SWE images, respectively.
An adaptive spatial fusion module is introduced to effectively combine the extracted features from these two modalities.
Additionally, an orthogonal regularization strategy is used to promote weight orthogonality, thereby mitigating feature redundancy.

In particular, the classification framework takes B-mode scan $X: \Omega \subset \mathbb{R}^{T \times H \times W \times C}$ and SWE scan $E: \Omega \subset \mathbb{R}^{T \times H \times W \times C}$ as input,
and outputs the corresponding prediction $P(Y|X,E;\theta)$ of csPCa,
where $Y$ denotes class labels and $\theta$ denotes the network's parameters.
$T$ indicates the frame number of the TRUS video, $H$ indicates frame height, $W$ indicates frame width and $C$ indicates channel number.

\subsection{Adaptive Spatial Fusion Module}
According to study~\cite{brown2015recent},
abnormal hypoechogenicity in B-mode images and abnormal stiff regions in SWE images are usually characterized as lesions for targeted biopsy.
Inspired by recent attention mechanisms for the feature aggregation~\cite{huang2023joint, wang2023a2fseg},
we introduce an adaptive spatial fusion module to measure the pixel-level contribution of these two modalities to final classification.

As shown in the right part of Fig.~\ref{fig:method},
the features $\mathbf{F}_X$ from B-mode and $\mathbf{F}_E$ from SWE are first added together.
Then the added features are concatenated with the modality-specific features and passed through the convolutional blocks to generate modality-specific spatial attention weights:
\begin{align}
\label{eq:fusion}
\mathbf{W}_X &= \sigma(
IN(
conv_X([1/2(\mathbf{F}_X + \mathbf{F}_E), \mathbf{F}_X])
)
), \\
\mathbf{W}_E &= \sigma(
IN(
conv_E([1/2(\mathbf{F}_X + \mathbf{F}_E), \mathbf{F}_E])
)
), 
\end{align}
where each convolutional block comprises a $1\times1\times1$ convolution with stride 2 followed by a instance norm $IN$ and a Sigmoid activation function $\sigma$.
Subsequently, the spatial attention weights are normalized by the Softmax function:
\begin{align}
\label{eq:softmax}
\mathbf{\hat{W}}_X &= \frac
{exp(\mathbf{W}_X)}
{exp(\mathbf{W}_X) + exp(\mathbf{W}_E)}, \\
\mathbf{\hat{W}}_E &= \frac
{exp(\mathbf{W}_E)}
{exp(\mathbf{W}_X) + exp(\mathbf{W}_E)}.
\end{align}
Then we perform voxel-wise multiplication of the spatial attention weights with the corresponding modality feature maps to obtain the adaptively fused feature maps $\mathbf{\hat{F}}$:
\begin{equation}
\label{eq:fushion}
\mathbf{\hat{F}} = \mathbf{\hat{W}}_X \cdot \mathbf{F}_X + \mathbf{\hat{W}}_E \cdot \mathbf{F}_E.
\end{equation}
The feature maps $\mathbf{\hat{F}}$ are then integrated back into the modality-specific network branches, as shown in the left part of Fig.~\ref{fig:method}.
With hierarchical feature fusion,
each branch is capable of capitalizing on the strengths of the other while preserving its unique characteristics.

\subsection{Orthogonal Regularization}
Orthogonal regularization~\cite{brock2016neural} is a regularization technique for convolutional neural networks,
which encourages weights to be orthogonal, thereby mitigating feature redundancy.
It simply enforces the weight matrices product $\mathbf{W}\mathbf{W}^\top$ to be close to the identity matrix $\mathbf{I}$,
where $\mathbf{W}$ denotes the weight matrix of the convolutional kernels.

Therefore, we employ the sum of the cross entropy loss and the orthogonal regularization loss to train the network:
\begin{equation}
\label{eq:loss}
\mathcal{L} = - \frac{1}{|\Omega|} \sum_{X \in \Omega}log(P(Y|X,E;\theta))
- \lambda \sum_{}(|\mathbf{W}\mathbf{W}^\top| - \mathbf{I}),
\end{equation}
where $\lambda$ is a weight coefficient.

\section{Experiments}
\label{sec:pagestyle}
\subsection{Dataset}
In this study, 
we conducted experiments on an in-house TRUS video dataset collected from the Cancer Center of Sun Yat-Sen University.
A total of 512 patients were included in this study and all patients received TRUS, SWE and prostate biopsy.
Among them, 346 patients had csPCa detected by biopsy while 166 patients were determined to be negative for csPCa.
We randomly employed 400 scans (271 with csPCa) for training,
and 112 scans (75 with csPCa) for testing.
Considering the computational cost, the TRUS videos were resized to $200 \times 144 \times 144 \times 3$.
The intensities were normalized to [0, 1].

\begin{figure}[t]
	\centering
	\includegraphics[width=0.91\linewidth]{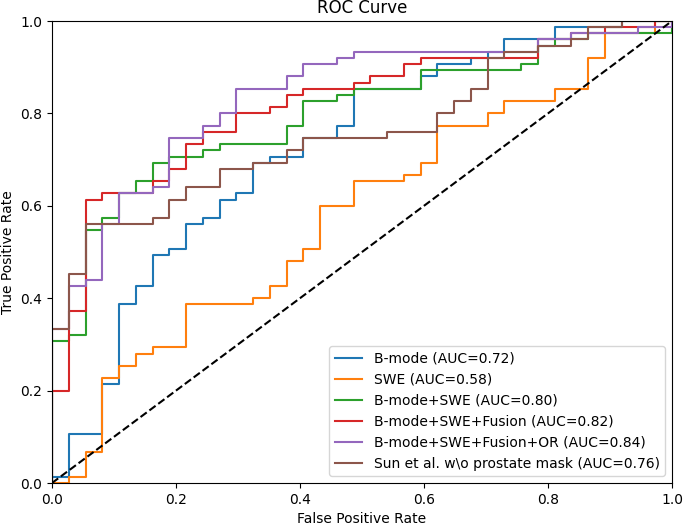}
	\caption{The receiver operating characteristic (ROC) curves of different methods on the testing set.}
	\label{fig:roc}
\end{figure}

\begin{table}[t]
	\centering                                     
	\caption{The comparison of different methods including one most relevant method~\cite{sun2023three} and several ablation models (best results are highlighted in bold).}
	\label{tab:result}
	\small
	\begin{tabular}{ccccccc}
		\toprule
		B-mode     & SWE        & Fusion     & OR         & AUC      & F1       & Acc \\
		\midrule
		\checkmark &            &            &            & 0.72     & 0.82     & 0.69\\
		& \checkmark &            &            & 0.58     & 0.81     & 0.68\\
		\checkmark & \checkmark &            &            & 0.80     & 0.82     & 0.72\\
		\checkmark & \checkmark & \checkmark &            & 0.82     & 0.83     & 0.74\\
		\checkmark & \checkmark & \checkmark & \checkmark 
		& \textbf{0.84}     & \textbf{0.86}     & \textbf{0.79}\\   
		\midrule
		\multicolumn{4}{c}{Sun~\textit{et al.}~\cite{sun2023three} w/o prostate mask}  
		& 0.76     & 0.80     & 0.66\\
		\bottomrule
	\end{tabular}
\end{table}

\begin{figure*}[t]
	\centering
	\includegraphics[width=\linewidth]{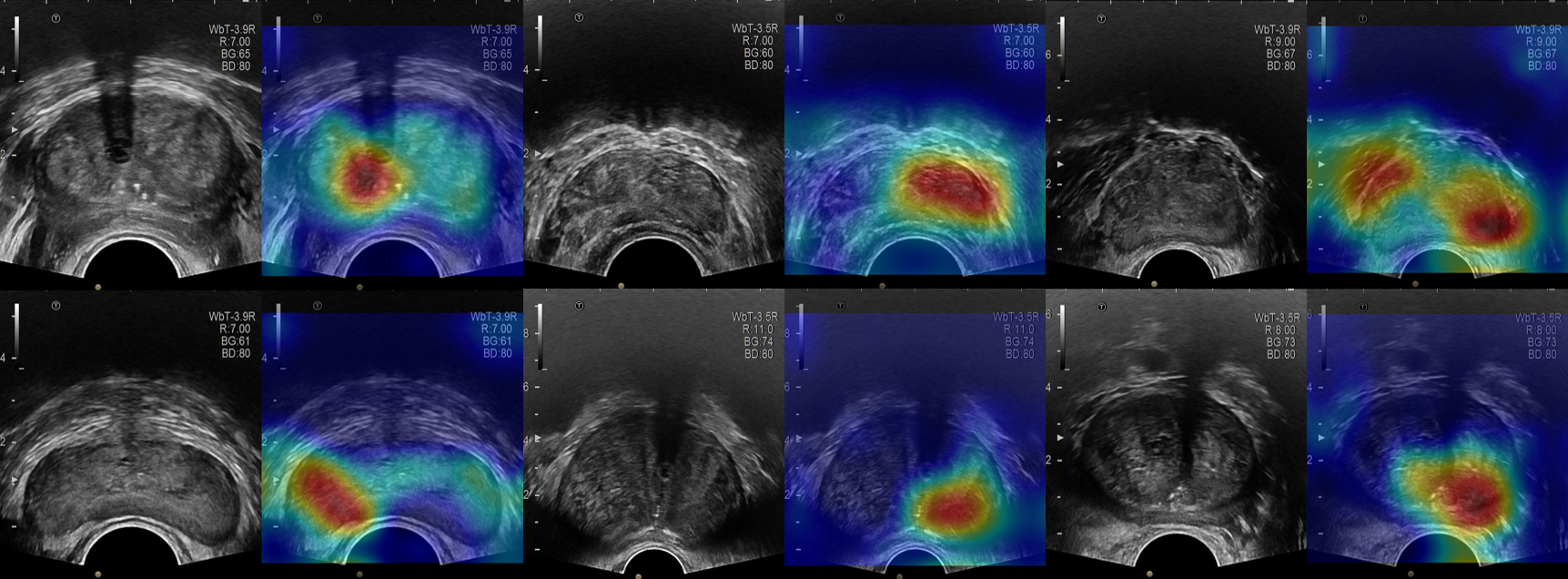}
	\caption{Six examples to show the TRUS frames and their corresponding heat maps generated using the class activation mapping (CAM). The regions highlighted in the heat maps contribute to the network's prediction of csPCa, and therefore may indicate the lesion areas associated with csPCa.}
	\label{fig:cam}
\end{figure*}

\subsection{Implementation Details}
Our proposed method was implemented in PyTorch, using 2 NVIDIA Tesla V100 GPUs with 40G memory.
We utilized two 3D ResNet-50 models as the backbone.
The whole network was optimized by a stochastic gradient descent (SGD) optimizer for 300 epochs, with an initial learning rate of 0.0001.
The ploy learning policy was employed to adjust the learning rate, $(1 - \text{epoch} / 200)^{0.9}$.
The training batch size was set to 2, consisting of a csPCa sample and a non-csPCa sample, which alleviated the class imbalance issue.
For the orthogonal regularization loss, we set $\lambda$ to 1e-5.
Our code is available at~\textit{https://github.com/2313595986/ProstateTRUS}.

\subsection{Quantitative and Qualitative Analysis}
\subsubsection{Quantitative Analysis}
We employed three typical classification metrics to evaluate csPCa identification performance,
including area under the ROC curve (AUC), F1-score (F1) and accuracy (Acc).
Higher scores for AUC, F1 and Acc indicate better classification performance.

To demonstrate the effectiveness of each component of our method,
we conducted ablation study on single modality (using only B-mode or SWE information),
and also performed ablation on the usage of adaptive spatial fusion module (Fusion) and orthogonal regularization (OR).
The numerical results are shown in Table~\ref{tab:result} and the receiver operating characteristic (ROC) curves are shown in Fig.~\ref{fig:roc}.
The proposed method achieved an AUC of 0.84, F1-score of 0.86, and accuracy of 0.79.
It can be observed that the concatenation of two modalities (i.e., early fusion) resulted in a obvious improvement on AUC,
compared to using B-mode only or SWE only.
Moreover, both adaptive spatial fusion module and orthogonal regularization further contributed to the improvement of classification accuracy.

We also compared our method with the study conducted by~\cite{sun2023three}.
To ensure a fair comparison, we excluded their segmentation module.
In the absence of prostate segmentation masks, our method outperformed theirs in terms of AUC, exhibiting an increase of 0.08.
This result highlights the efficacy of our method.

\subsubsection{Qualitative Analysis}
To determine the frames and specific locations within the TRUS video that contributed to the network's prediction of csPCa,
we employed the gradient-weighted class activation mapping 
(Grad-CAM)~\cite{selvaraju2017grad} to generate heat maps, as shown in Fig.~\ref{fig:cam}.
The regions highlighted in the heap maps may indicate the lesion areas associated with csPCa.
This finding suggests the potential feasibility of TRUS-guided targeted biopsy, by leveraging the CAM images generated using our method.

\section{Conclusion}
We introduce a simple yet effective method for the classification of clinically significant prostate cancer in TRUS videos.
The primary attribute is to employ the designed adaptive spatial fusion module to fully exploit complementary information from both B-mode and SWE images.
The experimental results on an in-house TRUS video dataset show the efficacy of the proposed method.
What's more, the visualized heat maps generated from the proposed method may provide valuable guidance for the localization of csPCa,
thus suggesting the potential feasibility of multi-modality TRUS-guided targeted biopsy.
Note that this study focuses on the classification of the csPCa.
One potential aspect to boost the classification performance is to leverage the prostate mask~\cite{sun2023three} by automated prostate segmentation~\cite{8698868}.
Another aspect could be using weakly-supervised methods~\cite{YANG2024122024, zhong2023simple} that leverage weak annotations to attain accurate lesion localization and therefore providing biopsy-guidance.

\hspace*{\fill}\\
\noindent\textbf{Compliance with Ethical Standards}\\
This research study was conducted retrospectively using medical ultrasound images and therefore the research received a waiver of approval from our institutional review board.

\hspace*{\fill}\\

\noindent\textbf{Conflicts of Interest}\\
The authors have no conflicts to disclose.

\bibliographystyle{IEEEbib}
\bibliography{refs}

\end{document}